\documentclass[aps,superscriptaddress,pra]{revtex4}
\usepackage{graphicx}

\newcommand{\non}{\nonumber}

\def\teps{t_{\epsilon}}
\def\la{\lambda}
\def\d2dot{\dot d_2}  
\def\lahat{\hat\lambda}
\def\eps{\epsilon}

\begin{document}

\title{A single saddle model for the $\beta$-relaxation 
in supercooled liquids}

\author{Andrea Cavagna}
\affiliation{Center for Statistical Mechanics and Complexity, INFM
Roma ``La Sapienza''}

\author{Irene Giardina}
\affiliation{Dipartimento di Fisica, Universit\`a di Roma ``La
Sapienza'', 00185 Roma, Italy}
\affiliation{Center for Statistical Mechanics and Complexity, INFM
Roma ``La Sapienza''}

\author{Tom\'as S.~Grigera}
\affiliation{Centro di Studi e Ricerche ``Enrico Fermi'', via
Panisperna 89/A, 00184 Roma, Italy}

\date{May 16, 2003}

\begin{abstract}
We analyze the relaxational dynamics of a system close to a saddle of
the potential energy function, within an harmonic approximation.  Our
main aim is to relate the topological properties of the saddle, as
encoded in its spectrum, to the dynamical behaviour of the system.  In
the context of the potential energy landscape approach, this
represents a first formal step to investigate the belief that the
dynamical slowing down at $T_c$ is related to the vanishing of the
number of negative modes found at the typical saddle point.  In our
analysis we keep the description as general as possible, using the
spectrum of the saddle as an input.  We prove the existence of a
time-scale $\teps$, which is uniquely determined by the spectrum, but
is not simply related to the fraction of negative eigenvalues.  The
mean square displacement develops a plateau of length $\teps$, such
that a two-step relaxation is obtained if $\teps$ diverges at $T_c$.
We analyze different spectral shapes and outline the conditions under
which the mean square displacement exhibits a dynamical scaling
identical to the $\beta$-relaxation regime of Mode Coupling Theory,
with power law approach to the plateau and power law divergence of
$t_\epsilon$ at $T_c$.
\end{abstract}

\maketitle

\section{Introduction}

In recent years much attention has been devoted in the field of
glass-forming liquids to the so-called 'energy landscape' paradigm.
The main background of this approach is the existence of a strong
relation between the topological properties of the potential energy
landscape (i.e.\ the potential energy as a function of the system's
coordinates) and the dynamical behaviour of the system.

Historically this idea goes back to Goldstein \cite{gold}, who, in
1969, argued how at low enough temperatures supercooled liquids spend
a long time in local minima of the potential energy, with rare
activated jumps among minima \cite{stillinger,zwanzig,glotzer}.
According to Goldstein this low-$T$ description breaks down at a
temperature $T_c$ above which activation is no longer the main
mechanism of diffusion.  This temperature $T_c$ has been subsequently
identified with the critical temperature where idealized Mode Coupling
Theory (MCT) locates a divergence of the relaxation time
\cite{bengtzelius, leutheusser,goetze}, supporting the view that $T_c$
marks a crossover from a low-$T$ activated hopping dynamics, to a
continuous flow dynamics at higher $T$, which is well described by MCT
\cite{angell,ullo,cummins,sokolov,doliwa}.

According to this interpretation, from a topological point of view, it
is clear that at low temperature the crucial role is played by the
minima of the potential energy function, since the system spends most
of the time close to them. This observation can be translated in
quantitative terms and various approaches have been elaborated to use
information available on the minima and their structure, to compute
(thermo)dynamical quantities \cite{stillinger,rahman,cotterill,
keyes-1}.

On the other hand, above $T_c$ the system is no longer confined around
a minimum, but rather 'flows' in configuration space. From a
topological point of view one may imagine that the system explores
regions of the configuration space which are rich in negative
modes. A confirmation of this perspective comes from the Instantaneous
Normal Modes (INM) approach \cite{rahman, cotterill}: the INM
spectrum, i.e. the average density of states of the Hessian matrix of
the potential energy, displays above $T_c$ a finite fraction of
imaginary modes, indicating that the system at equilibrium explores
regions of the landscape which have negative curvature
\cite{seeley,madan,keyes-1}.  There have been various attempts to use
the INM analysis to identify diffusive modes
\cite{cho,bembenek-1,sastry,sciortino,lanave}, but it still remains
unclear under what conditions this can be done, and how far the
procedure can be pushed.

From a theoretical point of view, a crucial question is whether also
above $T_c$ the behaviour of the system can be characterized by well
defined topological entities, as it happens at low temperature with
minima.  This point has been addressed in detail in mean-field
disordered systems in the context of spin-glasses
\cite{kurchan-1,cavagna-1,cavagna-2,franz,cavagna-2.2,thalmann}, as
well as in finite dimensional ordered models \cite{pettini}. Given the
numerous analogies between the phenomenology of certain spin-glass
models and that of structural glasses \cite{thirumalai}, and
exploiting the results achieved in that field, a new topological
interpretation of the dynamics above $T_c$ has been developed by
various authors. Within this scenario, above $T_c$ it is still
possible to describe the behaviour of the system in terms of well
defined topological entities. These are not minima, but rather
{\it saddle points} of the potential energy function, that is
stationary points with a non zero fraction of unstable directions.
More precisely, we can summarize this 'saddles interpretation' in a
few main points:
\begin{itemize}
\item[$\bullet$] For $T>T_c$ at equilibrium the system is 'close' to a
saddle point. The typical topological properties of this saddle point,
for example its energy $E$ and the number of unstable directions $K$,
depend on the temperature $T$.
\item[$\bullet$] Since $K\neq 0$ the system explores a region where
negative (non-confining) modes are available. Diffusion can then in
principle occur in two different ways: either via barrier hopping
along the positive eigenvalues, or exploiting the presence of these
free directions (see \cite{cavagna-3} for a discussion of this
point). When $K(T)$ is large this second mechanism will be faster than
the first one, and will reasonably prevail. However, upon lowering the
temperature $K(T)$ decreases and it will become less and less
efficient.
\item[$\bullet$] As the temperature approaches $T_c$ the typical
saddles become less unstable, and at $T_c$ they finally cease to have
negative modes and turn into minima.  Thus, as $T\to T_c$ the system
finds less and less free directions to diffuse and the second
mechanism of diffusion previously described will freeze.  If barriers
at $T_c$ are relevant, as numerical simulations indicate for fragile
systems \cite{grigera}, this implies a dramatic increase of the
relaxation time and a slowing down of the dynamics as $T_c$ is
approached. Around $T_c$ a crossover occurs to a different regime
where, being negative modes extremely rare, the main mechanism of
diffusion becomes barrier crossing and the Goldstein scenario is thus
recovered \cite{cavagna-3}.
\end{itemize}

This scenario has been tested in different ways and up to now various
indications in its favour have been gathered: Numerical experiments on
glass forming fragile models \cite{angelani, kurt, shah-1, grigera,
keyes-3, shah-2,wales,dawson} have shown that it is possible to
identify relevant saddles at a given temperature $T$ and classify them
through their index $K(T)$. Also, and more importantly, $K(T)\to 0$ as
$T\to T_c$, $T_c$ being the Mode Coupling temperature.  Besides,
analytical computations in simple models \cite{cavagna-4} show that
the system at equilibrium is truly close to a saddle according to a
rigorous well defined notion of distance in the configuration space.
As in numerical experiments, the index of the closest saddle goes to
zero as the critical temperature is approached.

Despite these encouraging results  the saddle scenario is not immune
of some criticisms. The results of numerical simulations 
\cite{kurt,grigera,angelani}
highly rely on numerical algorithms to locate the saddles, and the
sampling procedure has been deeply questioned (see for example \cite{wales}). 
Also the whole energy landscape approach, even if supported
by great part of the community, has recently received many objections
(see for example \cite{berthier} where a description of the
supercooled behaviour in terms of heterogeneities is preferred).
Besides, the saddle scenario  still remains
from a theoretical point of view rather vague and incomplete. In particular,
there are several questions which should be addressed, in our opinion, in a 
more formal and quantitative way:
\begin{itemize}
\item[i)] From an intuitive point of view and in the light of the
numerical results, it is reasonable that saddles do play a crucial
role in the dynamical behaviour of the system above $T_c$. However, is
this just a useful qualitative interpretation, or can we push this
approach further? In other words, is it possible to cast the 'saddle
dynamics' into a formal description?
\item[ii)] What is the precise relation between the vanishing of
negative modes and the increase of the relaxation time as $T\to T_c$?
In particular, what is the link between $K(T)$ and the relaxation time
$\tau$?
\item[iii)] Is $K(T)$ really the topological quantity most relevant to
the dynamics?
\item[iv)] Is a saddle description compatible with Mode Coupling
Theory?
\end{itemize}

In this paper we will address these questions in a formal way by
considering a very simple model of harmonic relaxation in a saddle.
In our analysis we only assume as starting hypotheses two facts which
have been numerically or analytically tested up to now: 1) the system
above $T_c$ is close to a saddle point with non zero index $K(T)$, and
2) the number of negative modes of this saddle goes to zero as $T\to
T_c$.  The first assumption will enable us to perform an harmonic
expansion of the potential energy around a saddle point, while the
second will be used as a constraint on the behaviour of the saddle
spectrum close to $T_c$.

Our aim is to find an explicit and quantitative link between the
topological properties of the energy landscape and the dynamical
behaviour of the system. The input of our model is then the spectrum
of the saddle, while our results concern the relation between this
spectrum and the relaxation time of the system.  We will leave the
shape of this spectrum unspecified as long as we can (apart from the
above-mentioned constraints) in order to make our statements as
general as possible. Only after will we discuss specific cases and
their physical relevance.  We will then show how even a simple
relaxational dynamics is able to provide precise answers to the
previous questions, and in certain cases, even approximately reproduce
the results of more complicated dynamical theories such as Mode
Coupling.

The paper is organized in the following way.  In Sec.~II we introduce
our model and the main quantity to be considered, namely the mean
square distance (MSD) (in configuration space) of the system from the
typical reference saddle point at temperature $T=(1+\epsilon)
T_c$. The MSD will be expressed in terms of the topological properties
of the saddle, more precisely its spectrum.  In Sec.~III we use the
Laplace method to find a relation between the asymptotic dynamical
behaviour of the MSD and the spectrum of the saddle point. In Sec.~IV
we analyze the consequences of our result, and discuss how the shape
of the spectrum may determine qualitatively and quantitatively the
long time dynamics. We look at different classes of spectra and focus
on the conditions under which a two-step relaxation with a diverging
time-scale is obtained. In Sec.~V we compare out results with the
prediction of the Mode Coupling Theory. Sec.~VI is devoted to our
conclusions.

\section{The harmonic saddle model}

Let us consider a glass forming system with $N$ particles and
potential energy $V({\bf r})$, where ${\bf r}= \{ r_i \}$ is the
(mass-weighted) vector of the particles positions and $i=1\dots N$ is
a particle index \cite{nota1}.  As explained in the Introduction, the
stationary points of $V$ are, from a topological point of view, the
quantities of interest to describe the behaviour of the system.

At equilibrium in the low temperature phase $T<T_c$, the system (i.e.\
its representative point $\underline r$) explores for most of the time
a region close to a minimum of $V$, only occasionally jumping via
barrier crossing into the basin of another minimum. This separation of
time scales between intra and inter basin motion has been exploited by
both the INM and the Inherent Structures \cite{stillinger} approaches
to compute dynamic (velocity autocorrelation) and thermodynamic
quantities (equilibrium energy, entropy etc.) via an harmonic
expansion of the potential energy around a reference configuration.
On the other hand, above $T_c$ the system is typically close to a
stationary point which is not a minimum, but a saddle with a certain
number of negative modes. One can wonder whether also in this case it
is possible to use this information to simplify the computations with
a reasonable approximation, as the harmonic one adopted in the low
temperature phase. If the relevant saddles have a large instability
index $K(T)$, the system very easily finds escape directions to flow
from one region of the phase to another one: even if the closest
stationary point will typically be a saddle with the same statistical
properties (index, energy etc.), it will not be the {\it same} saddle
even for relatively short times. However, we know from the numerical
and analytical works quoted in the Introduction, that when $T\to T_c$,
$K(T)\to 0$, thus close enough to $T_c$ the relevant saddles have a
very small instability index. Then, if at some time the system is
close to a given saddle it will take a long time before finding one of
the few available escape directions and flow away. In this case the
system remains close to the {\it same} saddle (we could say in its
'basin') for some time, and an harmonic expansion seems justified, at
least on time-scales shorter than the time when the saddle is
definitely left.

Given that, we shall now consider our system at a temperature
$T=(1+\epsilon)T_c$ where $\epsilon$ is a small parameter, and assume
that at time it is $t=0$ close to a saddle point whose coordinates are
$r_0$ in configuration space. We wish to examine the dynamical
behaviour of the system and to do that we resort to a simplified
treatment where we use an harmonic expansion of the potential $V$
around $r_0$.  Our starting point is then the following Langevin
equation:
\begin{equation}
\dot x_i = - \sum_{k=1}^N M_{ik} x_k + \xi_i(t) ,
\label{lange}
\end{equation}
where $x_i(t)=r_i(t)-r_i^0$ are the displacements from the saddle, the
matrix $M_{ik}$ is the second derivative of the potential $V$ at the
saddle point and $\xi$ is a $\delta$-correlated noise, $\langle
\xi_i(t)\xi_k(t')\rangle = 2T \delta_{ik}\delta(t-t')$ \cite{foot-3}.

This equation has formally a very simple form of the
Ornstein-Ulhenbeck type in $3 N$ dimensions. Of course it is not as
trivial as it may seem at a first sight in that the matrix $M_{ik}$ is
not known. This is not only a consequence of having left the potential
$V$ unspecified, since even when an explicit energy function is
assumed, still the exact position ${\bf r_0}$ of a typical saddle at
temperature $T$ is not determinable. This is the glassy nature of the
system we are considering: even if the Hamiltonian is deterministic,
as the temperature is lowered particles tend to arrange in disordered
configurations. Thus $M_{ik}$ has rather to be considered a disordered
matrix whose distribution is determined in a complicated way by the
equilibrium Boltzmann measure (see e.g.\
\cite{cavagna-4,euclidian}). Thus in the following the Hessian matrix
${\bf M}$ will be treated as a random variable whose statistical
properties are in principle accessible.  In particular, we will
express the dynamics in terms of the density of eigenvalues (spectrum)
of ${\bf M}$:
\begin{equation}
\rho(\la; r_0)=1/3N\sum_\alpha \delta(\la-\la_\alpha) \ ,
\end{equation}
where $\lambda_\alpha$ are the eigenvalues of $M_{ik}$. Of course,
also $\rho(\la; r_0)$ depends on the precise saddle where it is
evaluated, and it is thus a stochastic quantity, as the Hessian
is. However, its statistical properties are in general more easily
computed via numerical simulations \cite{angelani,kurt} or analytical
procedures \cite{noi-inm,grigera-2,stefano} and we have more information on
its typical behaviour.  For example, we know from previous works that
the instability degree of a typical saddle decreases as the critical
temperature is approached, which means that the spectrum evaluated in
a typical saddle has less and less negative modes. Consequently, we
will assume that the {\it average} spectrum $\rho(\la)$ depends
parametrically on $\epsilon\equiv (T-T_c)/T_c$, and that its negative
support vanishes for $T\to T_c$, i.e.\
\begin{equation}
\rho_\epsilon(0)\to 0 \qquad \mbox{for } \epsilon\to 0. 
\label{ass1}
\end{equation}

From eq.~(\ref{lange}) we can easily compute the mean square
displacement (MSD),
\begin{equation}
d_2(t)\equiv\frac{1}{3 N}\sum_{i=1}^{N} \sum_{a=1}^3\langle
x_i^a(t)^2\rangle = \frac{T}{3N}\sum_{\alpha=1}^{3N}
\frac{1-e^{-2\la_\alpha t}}{\la_\alpha}
= T \int_{-\infty}^{+\infty} d\lambda\
\rho_\epsilon(\la)\ \frac{1-e^{-2\lambda t}}{\la} ,
\label{d2}
\end{equation}
where we have only assumed that in the thermodynamic limit
fluctuations of the spectrum around its average value are negligible
(as indicated by numerical simulations), and we have therefore
substituted $\rho(\la;r_0)$ with $\rho_\epsilon(\la)$. Also, we have
considered as initial conditions $x_i(0)=0$, that is we have assumed
the system to start exactly on the top of the saddle. As we shall see,
the initial condition is not important as long as the system does not
start too far from the saddle (in Sec. III we will say exactly how
far). Indeed the additional term arising in equation (\ref{d2}) when
$x_i(0) \ne 0$ can be shown in this case to be 
exponentially small at large times.

This equation is the starting point of our analysis. All the physics
is clearly encoded in the behaviour of the spectrum, both in its shape
as a function of $\la$ and in its dependence on the temperature.  We
immediately see that if $\rho_\epsilon(\la)$ has some negative support
the integral will diverge for $t\to\infty$, meaning that the system
asymptotically leaves the saddle if there are some negative
eigenvalues, as expected. As already said, since we want our harmonic
approximation to be reasonable, we are working in the regime $\epsilon
\ll 1$, when the support of $\rho_\epsilon(\la)$ is almost entirely
positive. In this case the systems remains a long time close to the
saddle and we can study what happens in the large time limit, but {\it
before} the saddle is left.

\section{The asymptotic dynamics}

For mathematical convenience it is better to look at the time
derivative of $d_2(t,\epsilon)$,
\begin{equation}
\dot d_2(t,\epsilon) = 2T\int_{-\infty}^{+\infty} d\lambda\ 
\rho_\epsilon(\la)\ e^{-2 \lambda t} .
\label{d2dot}
\end{equation}
Since the MSD is a well defined physical quantity at any given time
$t$, this integral should converge for any value of $t$.  Convergence
for $\la\to-\infty$ requires that
\begin{equation}
\log\rho_\eps(\lambda) < 2 \lambda t \qquad \mbox{for } \lambda \to -\infty .
\end{equation}
If we exclude oscillating functions this implies a concavity condition
on $\rho(\la)$ (this is obvious graphically): it must exist a
$\lambda_0(\epsilon)$ such that
\begin{equation}
\frac{d^2}{d\lambda^2}\log \rho_\epsilon(\la) < 0 \qquad \mbox{for }
\lambda < \lambda_0(\epsilon) .
\label{condo}
\end{equation}
The support of the spectrum must be entirely positive for
$\epsilon=0$, and this implies that $\lambda_0(\epsilon)>0$ for
$\epsilon$ small enough. We can split the integral in (\ref{d2dot})
separating the domains $\la<\lambda_0$ and $\la >\lambda_0$. We
have
\begin{equation}
{\cal R} \equiv 2T \int_{\lambda_0}^{+\infty} d\lambda\
\rho_\epsilon(\la)\ e^{-2\lambda t} \leq \ \rho_{max} \
\frac{e^{-2\lambda_0 \, t}}{t}  ,
\label{resto}
\end{equation}
where $\rho_{max}$ is the maximum of $\rho_\epsilon(\la)$.  As we
shall see the remaining part of the integral in $\d2dot(t,\epsilon)$
is much larger than $\cal R$ for $t\gg 1$.  We can thus disregard
$\cal R$ and write
\[
\dot d_2(t,\epsilon)= 2T\int_{-\infty}^{\lambda_0} d\lambda\,
\rho_\epsilon(\la) e^{-2\lambda t} = 2T \int_{-\infty}^{\lambda_0}
d\lambda\, e^{-t S_\epsilon(\lambda,t)} ,
\]
with $S_\epsilon(\lambda,t)=2\lambda-\log \rho_\epsilon(\la)/t$.
Thus, only the left tail of the spectrum contributes to the behaviour
of $\d2dot(t,\epsilon)$.

To evaluate this integral in the regime $t\gg 1$ we can use the
Laplace (saddle point) method \cite{bender} to find:
\begin{equation}
\d2dot(t,\epsilon) = d_0 \, T \ \frac{e^{-t
S_\epsilon(\hat\lambda_\epsilon(t),t)}}
{\sqrt{t\,S_\epsilon''(\hat\lambda_\epsilon(t),t)}} ,
\label{tota}
\end{equation}
where $d_0$ is a constant independent of $t$ and $\epsilon$,
the prime indicates the derivative with respect to $\la$
and $\lahat_\epsilon(t)$ is the solution of the saddle point equation
$S_\epsilon'(\lahat_\epsilon,t)=0$, namely
\begin{equation}
\frac{\rho'_\epsilon(\lahat_\epsilon)}{\rho_\epsilon(\lahat_\epsilon)}
= 2t .
\label{saddle}
\end{equation}

From the convergence condition (\ref{condo}) we see that
$\rho_\epsilon'(\la)/\rho_\epsilon(\la)$ is a monotonically decreasing
function of $\la$, and therefore eq.~(\ref{saddle}) has a unique
solution $\lahat_\epsilon(t)$.  Besides, we see that
$\lahat_\epsilon(t)$ changes sign at a well defined time: if we define
\begin{equation}
t_\epsilon = \frac{1}{2}\ \frac{\rho_\epsilon'(0)}{\rho_\epsilon(0)},
\label{tempo}
\end{equation}
we have that
\begin{eqnarray}
\hat\lambda_\eps(t)>0 \qquad \mbox{for } t<t_\epsilon \\
\hat\lambda_\eps(t)<0 \qquad \mbox{for } t>t_\epsilon 
\end{eqnarray}
If we interpret the saddle value $\lahat(t)$ as the relevant mode at
time $t$, this result translates the intuitive fact that while at
short times the system does not realize the presence of escape
directions and 'relaxes' in the pseudo-basin around the saddle given
by the positive modes, as time increases it finally finds the unstable
modes and exploits them to flow away.  This argument already tells us
that $t_\epsilon$ represents a crucial dynamical time scale in our
problem. This can be appreciated in a more explicit and quantitative
way if we analyze in detail the behaviour of $\d2dot(t,\epsilon)$.

As we show in Appendix A, it is possible to rewrite~(\ref{tota}) as
\begin{equation}
\dot d_2(t,\epsilon) = D_0 \,T \sqrt{-\dot{\hat\lambda_\epsilon}(t)} \,
\exp\left(-2\int_{1}^t dt'\ \hat\lambda_\epsilon(t') \right) ,
\label{d2final} \label{D2FINAL}
\end{equation}
where $D_0$ is a constant. This equation shows that the sign of
$\lahat_\epsilon(t)$ is the key factor determining the asymptotic
behaviour of $\d2dot(t,\epsilon)$ and thus, depending on the value of 
$t$ compared to $\teps$, we find two time regimes:  

\begin{itemize}
\item[$\bullet$] {\bf Early time region ($1\ll t < t_{\epsilon}$):} In
this regime $\hat\lambda_\epsilon(t) > 0$ and thus $\dot
d_2(t,\epsilon)$ is a decreasing function of $t$, reaching its minimum
at $t=t_\epsilon$.  From (\ref{d2final}) we see that $\dot
d_2(t_\epsilon,\epsilon)\to 0$ for $t_\epsilon\to\infty$. This means
that if the time scale $t_\epsilon$ diverges at $T_c$, then
$d_2(t,\epsilon)$ develops a plateau, whose length is of order $\teps$
\cite{MSD}.  The value of $d_2$ at the plateau is given by
\begin{equation}
d_p(\epsilon) = T \int_{1/\teps}^\infty \!\frac{d\la}{\la} \,
\rho_\epsilon(\la).
\label{plato}
\end{equation} 
The demonstration of this expression is sketched in Appendix
B. Physically, it is rather intuitive from eq.~(\ref{d2}) if one
remembers that at $t=\teps$ only positive modes contribute, and makes
the additional approximation that those with $\lambda > 1/\teps$ have
reached their asymptotic value, while the rest do not contribute at
all.  Since the length of the plateau is proportional to $\teps$, it
is clear that the plateau itself is better defined the closer we are
to $T_c$. Note also that for large $\teps$, the expression for the
plateau is just the value of $d_2(t,\epsilon)$ which would be obtained
for a strictly positive spectrum by taking the limit $t\to\infty$ in
(\ref{d2}).  This means that for $1\ll t<\teps$ the system
``thermalizes'' in the saddle using only the positive eigenvalues, and
$d_2$ reaches the value it would have reached in a purely harmonic
well.  This justifies our initial condition $x_i(0)=0$: any other
choice with $d_2(t,\epsilon) \le d_p$ would have been the same.  The
average energy density is given by $E(t)= E_0 + \frac{1}{2}\,T\
[1-\d2dot(t,\epsilon)]$, where $E_0$ is the bare energy of the saddle.
Therefore, at the plateau, that is for $t\sim \teps$, the corrections
to the harmonic thermal energy $T/2$ are small if $\teps\ \gg 1$.

\item[$\bullet$] {\bf Late time region ($t \gg t_{\epsilon}$):} In
this regime we have that $\hat\lambda_\epsilon(t)<0$. Now the system
has finally found the unstable directions and its dynamics is ruled by
the escape from the saddle basin.  The integral in the exponential of
(\ref{d2final}) changes sign at a time $\tau$ defined by
$\int_{1}^\tau dt'\ \hat\lambda_\epsilon(t') = 0$.  Besides, in
Appendix C we show that the prefactor in the square root does not go
to zero exponentially for $t\to \infty$.  Thus, $\dot
d_2(t,\epsilon)\to\infty$ for $t\gg \tau$, corresponding to
$d_2(t,\epsilon)$ leaving the plateau, i.e.\ to the system leaving the
saddle. We could in principle identify $\tau$ as an 'escape' time and
consider it as a second relevant dynamical time scale.  However, as we
have stressed in the Introduction, our harmonic expansion is
meaningful as long as the system remains close to the saddle,
therefore it is not clear whether $\tau$ has a genuine physical
meaning or whether other terms in the potential expansion should
already be considered on these time scales.
\end{itemize}

We stress that these results are valid for a general
$\rho_\eps(\lambda)$, with minimal requirements on its behaviour
dictated by physical consistency.  To summarize, the most important
result is the existence of a relevant time-scale $\teps$, directly
expressed in terms of topological properties of the landscape (the
spectrum at the saddle point). This result answers one of the
questions discussed in the Introduction, namely the relation between a
topology and a relaxation time. Contrarily to the naive expectation,
the relaxation time $\teps$ is {\it not} naturally related to the
fraction of negative eigenvalues $k_\epsilon =
\int_{-\infty}^0\rho_\epsilon(\la)\,d\la$.  Rather, if there is a
degree of universality in saddle dynamics, the correct scaling
variable seems to be $\rho_\epsilon'(0)/\rho_\epsilon(0)$ (see
(eq. \ref{tempo})). A second general result is that the MSD exhibits a two
step relaxation close to the critical temperature, provided the
relevant time scale $t_\epsilon$ diverges at $T_c$. Given the
expression of $\teps$ and given that $\rho_\epsilon(0) \to 0$ for
$\epsilon \to 0$, it is likely that this is actually the most general
scenario for most of reasonable spectra shapes (even if there are
simple counter-examples, as the case $\rho_\epsilon(\lambda)=\epsilon
f(\lambda)$).

On the other hand, the way the system approaches and leaves the
plateau, which is encoded in expression (\ref{d2final}), depends on
the specific form of $\rho_\epsilon(\la)$.  Therefore, to extend
further our analysis we need at this point to specify the spectrum in
some detail. Thus in the next section we shall consider different
spectral shapes.

\section{Specific spectra}

Let us now consider some specific forms for the spectrum
$\rho(\lambda)$ and compute explicitly the behaviour of $d_2(t)$.
Since we are interested in the asymptotic behaviour we shall specify
the spectrum only in the left tail, that is for $\lambda<\lambda_0$.

\subsection{Exponential Case}

Let us consider
\begin{equation}
\rho_\eps(\lambda) \sim \rho_\eps(0) \exp (a \lambda) 
\end{equation}
where $\lim_{\eps \to 0} \rho_\eps(0)=0$ in order to ensure a positive
definite spectrum at $T_c$.  This spectrum is of the form 
$\rho_\eps(\lambda)=\eps \ f(\lambda)$ and, as noted in the previous section, it has 
a non-diverging time scale $t_\eps=a/2$.  
Besides, from expression (\ref{d2}) we see that
$d_2(t)$ diverges at a finite time $t=t_\eps$.  Equation (\ref{d2dot})
can be easily integrated and we get
\begin{equation}
\dot d_2(t,\epsilon)\sim 2 T \rho_\eps(0) \ {\rm erfc} (\lambda_0
(a-2t)) + {\cal R}.
\end{equation}
Therefore in this case we do not get any diverging time scale as a function
of $\epsilon$ nor a two step dynamics. Rather what happens is that
$d_2(t,\epsilon)$ diverges at the same time for any temperature above
$T_c$, but in a steeper and steeper way as $T_c$ is approached. 

\subsection{The Gaussian case}

The Gaussian tail is less trivial. In this case we consider
\begin{equation}
\rho_\eps(\lambda) \sim  e^{-\frac{b}{2}(\lambda-\bar\lambda_\eps)^2} ,
\label{gauss}
\end{equation}
with $\lim_{\eps\to 0}\bar\lambda_\eps^{-1}=0$. By applying the
Laplace method we find $\hat\lambda_\eps(t)=\bar\lambda_\eps-2t/b$ and
$t_\eps=b \bar\lambda_\eps /2$. Then $\dot d_2$ is given by
\begin{equation}
\dot d_2(t,\epsilon)\sim D_0 T \exp\left[
\frac{2t^2}{b}-2\bar\lambda_\eps t \right] ,
\label{gauss1}
\end{equation}
and 
\begin{equation}
d_2(t,\eps)\sim d_p(\eps)+ D_0 T e^{\bar\lambda_\eps \teps} 
\int_0^{t-\teps} e^{\frac{2 \tau^2}{b} } \, d\tau .
\label{gauss2}
\end{equation}
(The same expressions can of course be obtained by exactly solving the
Gaussian integral eq. (\ref{d2dot})). Therefore for a Gaussian tail we
get a two-step dynamics with a plateau and a diverging time-scale. The
plateau is approached exponentially, as we see from eq.~(\ref{gauss2}).

\subsection{The power law case}

This is actually the most interesting case, since it
gives predictions compatible with MCT. Besides, the $p$-spin spherical model 
which has been widely used as a mean-field description of glassy
physics and exactly obeys MCT equations, has a spectrum belonging to
this class (see Sec. V-C).

Let us consider
\begin{equation}
\rho_\epsilon(\la) = (\epsilon^\mu + \la)^\eta ,
\label{pow}
\end{equation}
with $\mu\eta >0$, such that $\rho_\epsilon(0)=\epsilon^{\mu\eta}\to
0$, for $\epsilon\to 0$.  From eq.~(\ref{saddle}) we have
$\lahat_\epsilon(t)= -\epsilon^\mu + \eta/2t$, while
eq.~(\ref{d2final}) gives
\begin{equation}
\d2dot(t,\epsilon)= D_0 T \frac{e^{2\epsilon^\mu t}}{t^{\eta+1}} .
\label{potenza}
\end{equation}
By integrating from $\teps$ (given by (\ref{tempo})) up to $t$, we find,
\begin{equation}
d_2(t,\epsilon) = d_p(\epsilon) + \epsilon^{\mu\eta}\ h(t/\teps), 
\qquad \teps= \frac{\eta}{2} \epsilon^{-\mu},
\label{superpotenza}
\end{equation}
where $h(x)$ is a scaling function obtained by the expansion of the
exponential in (\ref{potenza}). The approach to the plateau can be
found by noting that for $x\ll 1$ we have $h(x)\sim -1/x^\eta$, and
thus
\begin{equation}
d_2(t,\epsilon) - d_p(\epsilon) \sim - t^{-\eta} .
\label{lord}
\end{equation}

These three general shapes of the the spectrum give rise to different
dynamical behaviours. In particular, only a power spectrum gives rise
to a diverging time-scale {\it and} a power-law approach to the
plateau. This fact is particularly important when comparing the
predictions of this simple model with those of the MCT, as we shall do
in the next section.

To conclude, we note that a useful quantity to characterize the
approach to the plateau is the time-dependent effective exponent
\cite{crisanti},
\[
\nu_{\mathrm{eff}}(t,\epsilon)= 1 + \frac{d\log \dot
d_2(t,\epsilon)}{d\log t} = 1 + \frac{t}{2}\
\frac{\ddot{\hat\lambda}_\epsilon(t)}{\dot{\hat\lambda}_\epsilon(t)}
\,-\, 2t\,\hat\lambda_\epsilon(t) .
\]
If in some time regime the MSD has a power law dependence on time,
$d_2(t,\epsilon)= d_p \pm\, t^{\nu}$, this must show up as a constant
contribution to the effective exponent in some extended region of
time, that is $\nu_{\mathrm{eff}}(t)=\nu$.  If we apply this
definition to the previous cases, we see easily that the only one
exhibiting a constant region for $\nu_{\mathrm{eff}}(t)$ is the third
one (see Fig.~1), for which find $\nu_{\mathrm{eff}}(t)=
-\eta(1-t/\teps)$.

\begin{figure}
\includegraphics[clip,width=4 in]{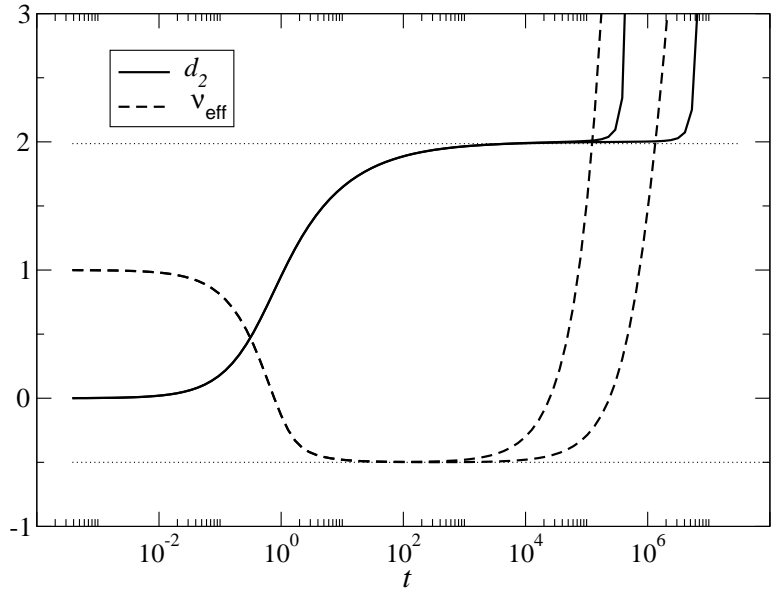}
\caption{The MSD (continuous line) and effective exponent (dashed
line) for a power law spectrum with $\eta=1/2$ and $\mu=1$, for
$\epsilon=10^{-5}$ and $\epsilon= 10^{-6}$. Dotted lines indicate the
plateau of $d_2$ from eq.~(\ref{plato}), and plateau of
$\nu_{\mathrm{eff}}$.}
\label{tong}
\end{figure}

\section{Comparison with MCT}

Mode Coupling Theory (MCT) \cite{bengtzelius, leutheusser,goetze} is a
dynamical approach which has been widely tested on a great range of
glass forming systems. It is not an exact theory, in that it makes
some assumptions on the memory functions entering the exact dynamical
equations, but it deals with the complete dynamics of the system and
takes into account in detail its structural properties.  For
structural fragile glasses it is now commonly accepted that MCT
represents an excellent description of the supercooled liquid system
for $T\geq T_c$ (see for ex. \cite{kob}).

Our single saddle model is, on the other hand, a much more simplified
approach and we cannot expect it to reproduce the whole set of MCT
results.  However, if we believe that close to $T_c$ the
main mechanism driving the dynamical slowing down is the vanishing of
the negative modes available to the system, our model should describe
reasonably the behaviour of the system in this temperature range and
should therefore reproduce at least the main MCT predictions.

\subsection{MCT predictions}

The MCT predicts the existence of a critical temperature $T_c$ where
the relaxation time of the system diverges. In real systems a real
dynamical  transition is not observed, 
however $T_c$ still represents a well defined relevant
temperature where a slowing down of the dynamics takes place and a
clear crossover to a different dynamical regime arrives.

A general prediction of MCT is the presence, as $T$ approaches $T_c$
from above, of a two-step relaxation of the relevant dynamical
quantities, such as, for example correlation, response functions and
the MSD.  More than this, MCT provides with detailed predictions about
the approach and escape from the plateau and the final relaxation to
the equilibrium asymptotic value. It distinguishes two dynamical
regimes, both of which become critical at the transition: the
so-called $\beta$-relaxation regime, which concerns the dynamics
around the plateau, and the $\alpha$-relaxation regime, which is
related to structural relaxation to the ergodic values.  Here, we will
be concerned with the $\beta$ relaxation, since, as already
underlined, our model is a reasonable approximation only until when
the plateau is left.

According to MCT, in the $\beta$-relaxation regime the correlation
function $C(t)$ has the following scaling form (as usual $t\gg1$),
\begin{equation}
C(t) = C_p + \epsilon^{1/2}\, g(t/\teps), \qquad
\teps=\epsilon^{-\frac{1}{2a}} .
\label{scaling}
\end{equation}
The correlation function has a plateau $C_p$ of length $\teps$, and
$\teps$ diverges as a power law for $\epsilon\to 0$. Moreover, MCT
predicts a power law approach to the plateau (early $\beta$ regime),
that is,
\begin{equation}
C(t) - C_p \sim t^{-a}, \qquad 1\ll t \ll \teps ,
\label{in}
\end{equation}
whereas for times larger than $\teps$ the correlation function leaves
the plateau as (late $\beta$ regime)
\begin{equation}
C(t) - C_p \sim - (t/t_0)^b , \qquad \teps<t\ll\tau ,
\label{out}
\end{equation}
where $t_0=\epsilon^{-\frac{a+b}{2ab}}$, and $\tau$ is the
$\alpha$-relaxation time.

For the supercooled system studied in \cite{kob} the value of the
exponent $a$ is $a=0.28$.

\subsection{Some general requirements}

Among the spectral shapes we have considered, only the power case
gives a scaling compatible with MCT. Looking at
equations~(\ref{superpotenza}) and~(\ref{lord}),
we conclude that the dynamical scaling of MCT is reproduced by spectra
of the form
\begin{equation}
\rho_\epsilon(\la) = (\epsilon^\mu + \la)^\eta, \qquad  \mu\eta=1/2 ,
\label{pow2}
\end{equation}
the predicted dynamical exponent being
$\alpha= \eta$.

Equation (\ref{pow2}), also implies a simple condition on the 
behaviour of the spectrum in $\lambda=0$: only spectra that behave as 
\begin{equation}
\rho_\eps(0) \sim \eps^{1/2}
\end{equation}
are compatible with MCT. This seems to be a quite strong topological 
requirement, very easy to check in specific cases.

Finally, we note that the important point is that the power law behaviour
is obeyed in the main range of the negative support as $T_c$ is
approached. The presence of small correcting tails around the lower
band edge $\lambda=-\eps^\mu$ 
is not important if they arise in a vanishing (as $\eps\to 0$)
region. Indeed, let us imagine that the power law behaviour
(\ref{pow2})  holds down to a certain value
$\bar\lambda>-\eps^\mu$. It is simple to show that the dynamical
behaviour predicted by a power law spectrum still holds up to
time-scales $t \sim \bar t$ with $\bar t=\rho_\eps '(\bar\lambda)
/2\rho_\eps(\bar\lambda)$, while for greater times the tail effect
starts dominating. Therefore if  $\teps/\bar t = 
(\eps^\mu+\bar\lambda)/\eps^\mu$  vanishes as $\eps \to 0$, the
dynamical scaling still holds in a diverging  time window.
In finite dimensional systems one usually gets small tails in the
spectrum where localized modes are concentrated. The previous
observation suggests that for a supercooled liquid, where we may
expect small tails to smoothen the spectrum close to the lower
band edge, only the extended modes representing the main negative
support are truly relevant for diffusion.

\subsection{The $p$-spin spherical model}

To perform a real comparison between MCT and our simplified harmonic
model we need to consider specific cases. In particular, an excellent
test would be a system where both approaches can be fully
applied. That is, we need a model where i) the MCT dynamical
phenomenology is reproduced and ii) information on the local topology
is analytically available.

The $p$-spin spherical model (PSM) \cite{crisanti} satisfies both
these requirements. Indeed, it can be shown that the {\it exact}
dynamical equations for this model have a MCT structure
\cite{bouchaud}, that is for the PSM the MCT is \emph{exact}
(contrarily to structural glasses where it represents an
approximation). Besides, for this model saddles have been shown
analytically to play a relevant role \cite{cavagna-4} and the spectrum
of a typical saddle at temperature $T=(1+\eps)T_c$ is exactly known
\cite{kurchan-2,kurchan-3}. Its left tail is
\begin{equation}
\rho_\epsilon(\la)=(\epsilon + \la)^{1/2}.
\end{equation}
Since in the PSM $d_2(t)=2\ [1-C(t)]$, the MCT scaling forms of
eqs.~(\ref{scaling}), (\ref{in}) and~(\ref{out}) directly apply also
to the mean-square displacement.  The exact resolution of the
dynamical MCT equations for $C(t)$ can be found in \cite{crisanti} and
gives $a_{PSM}\sim 0.4 $.

On the other hand the behaviour predicted by our harmonic model can be
simply inferred by the results of the previous sub-section with
$\mu=1$ and $\eta=1/2$. We get
\begin{equation}
d_2(t,\eps)=d_p(\eps)\sim + \eps^{1/2} \ h(t/\teps) ,	
\qquad \teps=\frac{1}{4 \eps},
\end{equation}
and a power law approach to the plateau:
\begin{equation}
d_2(t,\eps)-d_p(\eps)\sim - t^{-1/2}	.
\end{equation}	
This behaviour is consistent with a MCT scaling form with an exponent
$a=1/2$, to be compared with the exact value $a_\mathrm{PSM}\sim 0.4$.
Our conclusion is then that for the PSM the single saddle model
correctly reproduces the general dynamical scaling of the complete MCT
dynamics and the power law approach to the plateau of the early
$\beta$ regime.  The power law approach to the plateau can also be
seen by setting $t\ll\teps$ in the effective exponent. This gives,
$\nu_{\mathrm{eff}}(t,\epsilon)= -1/2 \, ( 1 - t/\teps)$.  From this
expression we see that for $t > \teps$ the model gives no power law
departure from the plateau (see Fig.~1) \cite{psm}.  This must not be
regarded as significant since the model itself looses validity at
large times (for this reason we have only focused on the early $\beta$
regime).

Unfortunately, our model does not exactly reproduce the exponent $a$
(or, in other terms, it does not give the correct form of the scaling
function $g(x)$ (\ref{scaling}), which is responsible for the value of
the exponent $a$).  Of course, this leaves us partly
unsatisfied. However, it is known that anharmonicities are very strong
in the PSM, and it is reasonable that they are already relevant close
to the saddle. If this is the case one should be able to show that
taking into account anharmonicities, for example via a perturbative
approach, modifies the exponent in the correct direction. We are now
working in this direction.

\section{Conclusions}

In this paper we have analyzed a simple model of relaxational dynamics
around an harmonic saddle. The physical problem we wanted to address
was the description of supercooled liquids close to the crossover
temperature $T_c$, and our main purpose was to model in a formal way
the connection between topological properties of the potential energy
and dynamical behaviour which is at the basis of the energy landscape
approach.

Despite the extreme simplicity of our model and the generality of our
assumptions our results are not trivial. We have obtained a
general expression which relates the spectrum of the saddle and a
relevant time-scale, and we have outlined the conditions under which
such a time-scale diverges giving rise to a two-step dynamics with a
well defined plateau at $T_c$. This time-scale is related to unstable
modes, however it is not proportional to the instability index, as
naive expectation would have suggested. Rather, it is related to the
behaviour of the spectrum close to zero: since negative modes tend to
disappear as $T_c$ is approached it is the way they turn into soft
modes that determines the long time dynamics.

The precise way the system relaxes to the plateau depends on the shape
of the saddle spectrum. We have analyzed different spectral shapes and
we have shown that power law spectra give rise to the same dynamical
scaling as the one predicted by MCT in the $\beta$-regime. For this
kind of spectra then our model seems to be consistent with MCT.

Beside these general qualitative predictions we have also looked at
the specific case of the $p$-spherical model, where both the spectrum
and the dynamics can be analytically computed. Here the saddle model
reproduces the dynamical scaling predicted by the MCT, which is in
this case exact.  The $p$-spin spherical model is interesting because
it represents a concrete model where the two approaches, MCT and
single-saddle relaxation, can be independently performed and
ultimately compared. Our analysis indicates that, despite its brutal
disregarding of anharmonic contributions, the single saddle model
already reproduces the main features of the complete exact dynamics,
namely the dynamical scaling and power law approach to the
plateau. Our interpretation of this result has been sketched in the
Introduction: the presence of the plateau and the slowing down of the
dynamics described by the MCT equations are mainly due to the
vanishing of the negative modes close to the transition temperature
$T_c$, and these effects are already taken into account at the
harmonic level in the single saddle model.  In the PSM we have
explicitly demonstrated this, however we expect the same to hold
whenever MCT is a good description of the problem, as for structural
fragile glasses. To support this conclusion we should in principle
proceed as we did for the PSM: compute the dynamical quantities using
the single saddle model and compare them with the MCT predictions. The
problem is that, contrarily to the $p$-spin case, the spectrum of a
typical saddle is not in general known for models of glass forming
systems. We may expect that numerical simulations or analytic
computations will provide this topological information and our
analysis be completed in the future.  On the other hand, there is
something important we have already done at this stage: we have
outlined under what general conditions for $\rho(\lambda)$ our single
saddle model is compatible with the MCT predictions. If the spectra of
real glass systems will turn out to satisfy these conditions 
we will then have very strong support for
our arguments.

\begin{acknowledgments}
We acknowledge rewarding discussions with S.~Ciliberti, G.~Parisi,
F.~Ricci-Tersenghi and P. Verrocchio.
\end{acknowledgments}

\appendix

\section{Computation of the MSD}

Here we show how to derive expression (\ref{d2final}) for the MSD.
For convenience we omit in the following the sub-index $\epsilon$. We
remind that the dot indicates a derivative with respect to time, while
the prime a derivative with respect to $\lambda$.  Let us call
$N(t)=e^{-tS(\hat\lambda)}$.  By deriving $N(t)$ with respect to time
the numerator of eq. (\ref{tota}) and using the saddle point equation
(\ref{saddle}) we obtain
\begin{equation}
\dot N(t)=-2 \hat\lambda(t) \ N(t),
\end{equation}
and thus
\begin{equation}
N(t)=N(t_0)\exp\left(-2\int_{t_0}^t dt'\ \hat\lambda(t') \right).
\end{equation}
For what concerns the denominator of (\ref{tota}) we have to evaluate
\begin{equation}
tS''(\hat\lambda)=\left(
\frac{\rho'^2}{\rho^2}-\frac{\rho''}{\rho}\right)=
t^2-\frac{\rho''(\hat\lambda)}{\rho(\hat\lambda)},
\end{equation}
where we have used again the saddle point equation. We define
$f(t)=\rho(\hat\lambda(t))$. Derivatives of $f(t)$ with respect to
time give
\begin{eqnarray}
\dot f(t)&=& t \dot{\hat\lambda}  f(t), \\
\frac{1}{f(t)} \ddot f(t) &=& 
\frac{\rho''(\hat\lambda)}{\rho(\hat\lambda)}  \dot{\hat\lambda}^2 +
t\ \ddot{\hat\lambda}, \\
\frac{1}{f(t)} \ddot f(t) &=& 
\dot{\hat\lambda} + t\ddot{\hat\lambda}+ t^2 \dot{\hat\lambda}^2.
\end{eqnarray}
Putting together these relations we have
\begin{equation}
tS''(\hat\lambda) = -\frac{1}{\dot{\hat\lambda}}.
\end{equation}
So finally eq.(\ref{tota}) becomes
\begin{equation}
\dot d_2(t,\epsilon) = d_0  \sqrt{-\dot{\hat\lambda}(t)}
\exp\left(-2\int_{1}^t dt'\ \hat\lambda(t') \right), \qquad t \gg 1,
\label{d2app}
\end{equation}
where we have fixed $t_0=1$ for convenience, and $d_0=N(1)$. 

We recall that this formula is valid for $t\gg 1$, and on the
assumption that the integral for $\lambda>\lambda_0$ is
sub-dominant. We can see from (\ref{d2}) that this is actually
true. Indeed, since the saddle point solution $\hat\lambda$ is a
decreasing function of time, it will surely become smaller that
$\lambda_0$, for $t$ large enough, therefore
\begin{equation}
\int_{1}^t dt'\ \hat\lambda(t') \ll \lambda_0 t
\end{equation}
at large times and ${\cal R}(t)$ is exponentially sub-dominant with
respect to (\ref{d2app}).

\section{The plateau of the MSD}

We now want to determine the value of the plateau.  First, we need to
appropriately define what we mean with plateau.  We have seen that
$\dot{d}_2(t)$ has a minimum for $t\sim t_\epsilon$.  This means that
$d_2(t_\epsilon)$ is as close as possible to the limiting value where
$d_2(t)$ would go if there were no negative eigenvalues.  Thus, we may
define an effective plateau for $\epsilon \neq 0$ (but very small) as
\begin{equation}
q_\epsilon = d_2(t_\epsilon).
\label{plat}
\end{equation}
In this way for $\epsilon \to 0$ $q_\epsilon$ approaches a well
defined value $q(T_c)$ related to relaxation in a harmonic well.  We
now show that
\begin{equation}
q_\epsilon = \int_{1/t_\epsilon}^{\infty}
\frac{\rho(\lambda)}{\lambda} + {\rm O} \left [ \rho(1/t_\epsilon)
\right ] + {\rm O}[\rho(-1/t_\epsilon) ],
\label{platfin}
\end{equation}
and thus 
\begin{equation}
q_\epsilon \to q(T_c)=\int^{\infty}_{0}\frac{\rho(\lambda)}{\lambda}
\qquad \mbox{for } \epsilon \to 0,
\end{equation}
as expected.  To prove (\ref{platfin}) we proceed as follows. We split
$d_2(t)$ into different parts
\begin{eqnarray}
d_2(t) &=& \int_{-\infty}^{\infty} \!\! d\lambda\, \rho(\lambda) 
\frac{1-e^{-2\lambda t}}{\lambda} \equiv
A+B+C+D+E+F=  \non \\
&& \int_{1/\teps}^{\infty} \frac{\rho(\lambda)}{\lambda} \, d\lambda
-\int_{1/\teps}^{\infty} \frac{\rho(\lambda)e^{-2\lambda t}}{\lambda}
\,d\lambda + \int_{0}^{1/\teps} \rho(\lambda) \frac{1-e^{-2\lambda
t}}{\lambda} \, d\lambda + \non \\
&& \int^{-1/\teps}_{-\infty} \frac{\rho(\lambda)}{\lambda} \, d\lambda
-\int^{-1/\teps}_{-\infty} \frac{\rho(\lambda)e^{-\lambda t}}{\lambda} \,
d\lambda + \int^{0}_{-1/t_\eps} \rho(\lambda) \frac{1-e^{-2\lambda
t}}{\lambda} \, d\lambda ,
\label{mink}
\end{eqnarray}
and we evaluate the different integrals.
By using the saddle point method for $t>t_{\eps}$ we find
\begin{equation}
B \sim -\frac{t}{\teps} \rho(1/\teps) e^{-2 t/\teps} \qquad
\Longrightarrow \qquad B\sim \rho(1/t_\eps) \qquad \mbox{for } t\sim
\teps .
\end{equation}
C is easy:
\begin{equation}
C \leq \rho(1/t_\eps) \int_0^{t/t_\eps} \!\! dx\, \frac{1-e^{-2 x}}{x}
\qquad \Longrightarrow \qquad C\sim c \rho(1/t_\eps) \qquad \mbox{for
} t\sim t_\eps,
\end{equation}
where $c$ is a constant prefactor.  To evaluate D we note that to
ensure the convergence of $d_2(t)$ the function
$f(\lambda)=\rho(\lambda) e^{-2 \lambda t}$ should go to zero for
$\lambda \to -\infty$. We can thus estimate D with the saddle point
method:
\begin{equation}
D=\int_{-\infty}^{-1/t_\eps} f(\lambda)\frac{e^{2 \lambda t}}{\lambda}
\,d\lambda \sim \frac{t_\eps}{t}f(-1/t_\eps) e^{-2 t/t_{\eps}} \qquad
\Longrightarrow \qquad D\sim \rho(-1/t_\eps) \qquad \mbox{for } 
t\sim t_\eps.
\end{equation}
The integral E is trickier. Once again we use the saddle point method.
The procedure is the same as the one used for ${\dot d}_2(t)$.  Since
the factor $1/\lambda$ is algebraic it can be discarded in the
evaluation of the maximum so that the maximum of the integrand still
occurs for a value $\hat\lambda(t)$ as given by equation
(\ref{saddle}).  But, by definition, $\hat\lambda(t_\eps)=0$. So for
$t$ close enough to $t_\eps$ the real maximum lies outside the
integration range and the integral is dominated by the upper
integration limit. Thus we have
\begin{equation}
E \sim \frac{-t_\eps}{-t+\frac{\rho'(-1/t_\eps)}{\rho(-1/t_\eps)} }
e^{2t/t_{\eps}} \rho(-1/t_\eps) \qquad \Longrightarrow \qquad
E \sim \rho(-1/t_\eps) \qquad \mbox{for } t\sim t_\eps.
\end{equation}
Finally, exactly as for C we have
\begin{equation}
F \leq \rho(-1/t_\eps) \int^0_{-t/t_\eps} \!\! dx \,
\frac{1-e^{-x}}{x} \qquad \Longrightarrow \qquad F\sim 
\rho(-1/t_\eps) \qquad \mbox{for } t\sim t_\eps .
\end{equation}

\section{The prefactor of eq.~(\protect\ref{d2final})}

We now argue that the prefactor of the square root of
eq.~(\ref{d2final}) does not go to zero exponentially for
$t\to\infty$, at least for physical spectra shapes.  Let us suppose
that for $t\to \infty$ we have
\begin{equation}
-\dot{\hat\lambda}(t) \le A e^{-at}.
\label{ass}
\end{equation}
This  implies that
\begin{equation}
\hat\lambda(t)\to \lambda_\infty \qquad t\to \infty,
\end{equation}
that is, if we define  $g(\lambda)=\frac{\rho'(\lambda)}{2 \rho(\lambda)}$,
\begin{equation}
g(\lambda)\to\infty \qquad \lambda\to\lambda_\infty,
\end{equation}
with $\lambda_\infty <0$.  This is what happens when $\rho$ has a cut
in the left tail (semicircle).  Condition (\ref{ass}) implies (we
integrate between $t$ and $\infty$)
\begin{equation}
\hat\lambda(t) \le \lambda_\infty + A/a e^{-at}.
\end{equation}
Given that $g(\lambda)$ is a decreasing function of $\lambda$, using
the saddle point equation we find
\begin{equation}
2 t= g(\lahat) \ge g\left(\lambda_\infty  + A/a  e^{-at}\right).
\end{equation}
Calling $x\equiv \lambda_\infty + A/a e^{-at}$, and for
$x\sim\lambda_\infty$, this relation gives
\begin{equation}
g(x)\le - \frac{1}{a}\log(x-\lambda_\infty).
\end{equation}
If we now recall that $g(x)=\rho'(x)/\rho(x)$, and we integrate
between $\lambda$ and $0$ and take the limit $\lambda\to
\lambda_\infty$, we find
\begin{equation}
\rho(\lambda_\infty) > 0.
\end{equation}
Thus, if equation~(\ref{ass}) holds, the spectrum must be nonzero at
the left cut (the converse is not true). Therefore, if we discard this
unphysical case, we are sure that for $t\to\infty$ the Gaussian
fluctuations do not go to zero exponentially.  It is possible that for
spectra of the kind we exclude, the exponential factor increases
faster than a simple exponential and therefore still kills the
prefactor contribution, but we have not been able to prove under what
conditions this happens.


\begin{thebibliography}{22}
\bibitem{nota1}
The particle position $r_i$ is still a vector in $3$ dimesions, 
i.e.  $x_i= \{x_i^a\}$ with $ a=1,\dots 3$,
however we will indicate it with a roman variable in order to keep the 
notation simple.
\bibitem{foot-3} We assume that the system is at the top of 
the saddle at $t=0$, i.e. $x_i(0)=0 \  \forall i$. This
will be justified later.
\bibitem{foot-8}
In the PSM we have simply  $d_2(t)=2[1-C(t)]$.
\bibitem{psm}
In fact, the effective exponent in the PSM shows that even for
$\teps\gg 1$ there is no extended time region where
$\nu_{\mathrm{eff}}=b$ (Fig.6 of \cite{crisanti}), and thus the
departure from the plateau is not really power law in the PSM.
\end{thebibliography}
\end{document}